\documentclass[aps,pre,reprint]{revtex4-1}
\usepackage{amsmath}
\usepackage{graphicx}
\usepackage{color}
\usepackage{amssymb}
\usepackage{wasysym}
\usepackage{soul}
\input{epsf}

\begin{document}

\bibliographystyle{prsty}

\title{Free Minimization of the Fundamental Measure Theory Functional: Freezing of Parallel Hard Squares and Cubes}

\author{S. Belli$^1$}
\email{e-mail: s.belli@uu.nl}
\author{M. Dijkstra$^2$}
\author{R. van Roij$^1$}

%\affiliation{$^1$Institute for Theoretical Physics, Utrecht University, Leuvenlaan 4, 3584 CE Utrecht, The Netherlands \\
%$^2$Soft Condensed Matter Group, Debye Institute for NanoMaterials Science, Utrecht University, Princetonplein 5, 3584 CC Utrecht, The Netherlands}

\affiliation{$^1$ Institute for Theoretical Physics, Utrecht University, Leuvenlaan 4, 3584 CE Utrecht, The Netherlands}
\affiliation{$^2$ Soft Condensed Matter Group, Debye Institute for NanoMaterials Science, Utrecht University, Princetonplein 5, 3584 CC Utrecht, The Netherlands}

\begin{abstract}

Due to remarkable advances in colloid synthesis techniques, systems of squares and cubes, once an academic abstraction for theorists and simulators, are nowadays an experimental reality. By means of a free minimization of the free-energy functional, we apply Fundamental Measure Theory to analyze the phase behavior of parallel hard squares and hard cubes. We compare our results with those obtained by the traditional approach based on the Gaussian parameterization, finding small deviations and good overall agreement between the two methods. For hard squares our predictions feature at intermediate packing fraction a smectic phase, which is however expected to be unstable due to thermal fluctuations. This implies that for hard squares the theory predicts either a vacancy-rich second-order transition or a vacancy-poor weakly first-order phase transition at higher density. In accordance with previous studies, a second-order transition with a high vacancy concentration is predicted for hard cubes.

\end{abstract}

\maketitle

\section{Introduction}

Hard spheres represent the simplest and most versatile model for the description of molecular and colloidal many-particles systems. This statement is particularly true since 1957, when Wood and Jacobson \cite{wood} and Alder and Wainwright \cite{alder} demonstrated that hard spheres undergo a fluid-to-crystal transition, and therefore that hard interactions alone can account for freezing. 

Systems of hard cubes, on the other hand, were considered as mere toy models until only a few years ago. The reason for this is evident: no molecule or macromolecular aggregate found in nature is known to be reasonably approximated by this shape. However, the interaction between parallel hard cubes is the second-simplest hard interaction one can imagine after that between hard spheres. Its simplicity made this model a perfect object of study for theory and simulation. 

Early studies on the equation of state of parallel hard squares ($D=2$ dimensions) and cubes ($D=3$) date back to the dawn of computer simulation in the 1950s \cite{zwanzig, hoover1}. Soon after, the question regarding the high-packing phase behavior of the models arose. For parallel hard squares, a transition from the fluid to a square-lattice crystal (with quasi-long-range order) was found \cite{beyerlein1, rudd}, but its character, whether continuous or discontinuous, has been a matter of debate ever since \cite{carlier, ree, frisch, hoover2}. Conversely, the stability of a ``brick-wall'' smectic phase with one-dimensional ordering in rows (or columns) was suggested to exist for parallel hard squares, but the stability of this peculiar state was soon ruled out \cite{beyerlein2}. Similarly, parallel hard cubes manifest a fluid-to-crystal transition with a well-established second-order character \cite{jagla} and no stable phase with lower translational symmetry than the crystal is expected \cite{groh}.

In the mid-1990s, while the interest of the liquid-state community was focusing on mixtures of hard spheres, hard cubes were rediscovered. By means of computer simulation Dijkstra et al. showed evidence of a demixing transition in a binary system of parallel hard cubes on a lattice, thus demonstrating that additive hard interactions can induce an entropy-driven fluid-fluid phase separation \cite{dijkstra1,dijkstra2}. These results motivated Cuesta and Mart\'{\i}nez-Rat\'{o}n to face the problem by means of density functional theory. Following the pioneering approach developed by Rosenfeld for hard spheres \cite{rosenfeld}, they developed a Fundamental Measure Theory (FMT) \cite{fmtreview} formalism aimed at describing both the homogeneous and inhomogeneous phase behavior of mixtures of squares and cubes \cite{cuesta, mraton}. 

Since the early work on hard squares and cubes, the progress in colloidal particles synthesis has been enormous. In particular, colloidal suspensions of micron-sized cubes \cite{rossi} and quasi-two-dimensional square platelets \cite{zhao} have been recently produced and analyzed. These experimental advances led to a renewed interest in the model, and at present more complex aspects like the role of orientational degrees of freedom, the addition of dipolar interactions, the roundedness of the shape and the effect of vacancies in the freezing mechanism constitute objects of intense research \cite{smallenburg, zhang, batten,avendano,ni,marechal}. Far from being a toy model or a mere academic exercise, squares and cubes have therefore gained a key role as model systems of non-spherical colloidal particles.   

Besides the development of new theories \cite{hansengoos, hansengoos2}, the increasing attention towards the self-assembly of non-spherical particles requires a detailed analysis of the capabilities of the existing ones. The aim of this paper is to reinvestigate the prediction of Fundamental Measure Theory as formulated in Ref. \cite{cuesta} for the phase behavior of parallel hard squares and cubes. The focus of our attention points to the freezing transition and the structure of the high-density inhomogeneous phases. In particular, by exploiting present-day computer power we improve previous analyses on the subject by performing a {\em free minimization} of the density functional, and compare our results with those obtained by means of the widely applied Gaussian parameterization of the single-particle density. We observe good overall agreement between the two methods, the main drawbacks of the Gaussian parameterization being (i) a systematic albeit small underestimation of the equilibrium vacancy 
concentration in the crystal and (ii) the lack of anisotropy of the crystal density peaks at high enough density. Furthermore, to the best of our knowledge this work constitutes the first density-functional theory study of the hard-square system. We show that Fundamental Measure Theory surprisingly predicts a smectic phase absent in computer simulation and suggests that, in analogy with the hard-cube system, vacancies can play a crucial role in the freezing transition.    

\section{Density Functional Theory}

The density-functional theory route to the equilibrium properties of a many-body system consists of expressing the intrinsic Helmholtz free energy $\mathcal{F}$ as a functional of the single-particle density $\rho(\mathbf{r})$ \cite{evans}. When considering a system composed of a single species of particles having only translational (and no rotational) degrees of freedom in $D$ dimensions, the free-energy functional reads

\begin{equation}
\label{eq1}
\beta \mathcal{F}[\rho] = \int d^D \mathbf{r} \, \rho(\mathbf{r}) \Bigl \{ \log[\rho(\mathbf{r}) \Lambda^3]-1 \Bigr \} + \beta \mathcal{F}^{\mathrm{exc}}[\rho],
\end{equation}
where $\mathbf{r}$ is a $D$-dimensional vector, $\beta=(k_B T)^{-1}$ is the inverse temperature in units of the Boltzmann constant, $\Lambda$ the thermal wavelength and the integrals are performed over the ($D$-dimensional) volume $V$ occupied by the system. The first term in the right-hand side of Eq. (\ref{eq1}) denotes the ideal-gas contribution, while the second describes the excess contribution due to particle-particle interactions.

\subsection{Fundamental Measure Theory (FMT)}

The excess free-energy functional $\mathcal{F}^{\mathrm{exc}}$ in Eq. (\ref{eq1}) is the non-trivial element of the theory: it contains the free-energy dependence on the inter-particle interactions and it can not be calculated exactly in general. 

Various methods to systematically estimate this functional dependence have been developed. For hard spheres the undoubtedly most successful approach is that of Fundamental Measure Theory (FMT). According to FMT, the excess free energy is written as

\begin{equation}
\label{eq3}
\beta \mathcal{F}^{\mathrm{exc}}[\rho] = \int d^D \mathbf{r} \, \Phi^{(D)}\bigl(\{n_{\alpha}(\mathbf{r})\}\bigr),
\end{equation}
where $\{n_{\alpha}(\mathbf{r})\}$ is a set of weighted densities, labeled by $\alpha$, obtained as convolutions between the single-particle density and a set of corresponding weight functions $\mathrm{w}_{\alpha}(\mathbf{r})$, 

\begin{equation}
\label{eq4}
n_{\alpha}(\mathbf{r}) = \int d^D \mathbf{r}' \, \rho(\mathbf{r}') \mathrm{w}_{\alpha}(\mathbf{r}-\mathbf{r}').  
\end{equation}
The functional dependence of $\Phi^{(D)}\bigl(\{n_{\alpha}\}\bigr)$ is determined by extrapolating from known limiting cases, such as the homogeneous bulk equation of state, the low-density second-virial behavior, and the dimensional crossover to highly confined conditions \cite{fmtreview}.

For hard parallel squares ($D=2$) and cubes ($D=3$) with side $\sigma$, the FMT functional was determined by Cuesta and Mart\'{\i}nez-Rat\'{o}n in Ref. \cite{cuesta}. In what follows, we report the explicit expression of $\Phi^{(D)}\bigl(\{n_{\alpha}\}\bigr)$ for the single-component case. Following Ref. \cite{cuesta} we introduce the auxiliary functions

\begin{equation}
\label{eq5}
\tau(x) = \Theta \biggl(\frac{\sigma}{2} -|x|\biggr), \, \, \, \zeta(x) = \frac{1}{2} \delta\biggl(\frac{\sigma}{2} -|x|\biggr), 
\end{equation}
defined for $x \in \mathbb{R}$.

For parallel squares the weight functions are
 
\begin{subequations}
\begin{flalign}
\label{eq6}
& \mathrm{w}_0(\mathbf{r}) = \zeta(x) \zeta(y); \\
& \mathbf{w}_1(\mathbf{r}) = \bigl( \zeta(x) \tau(y), \tau(x) \zeta(y) \bigr); \\
& \mathrm{w}_2(\mathbf{r}) = \tau(x) \tau(y), 
\end{flalign}
\end{subequations}
where we note that $\mathbf{w}_1$ has a vector character. The functional dependence of the excess free energy of parallel cubes is given by

\begin{equation}
\label{eq7}
\Phi^{(2)} = -n_0 \log \bigl(1- n_2 \bigr) + \frac{n_1^{(x)} n_1^{(y)}}{1- n_2}.
\end{equation}
For parallel cubes the weight functions are

\begin{subequations}
\begin{flalign}
\label{eq8}
& \mathrm{w}_0(\mathbf{r}) = \zeta(x) \zeta(y) \zeta(z);\\
& \mathbf{w}_1(\mathbf{r}) = \bigl( \tau(x) \zeta(y) \zeta(z), \zeta(x) \tau(y) \zeta(z), \zeta(x) \zeta(y) \tau(z)); \\
& \mathbf{w}_2(\mathbf{r}) = \bigl( \zeta(x) \tau(y) \tau(z), \tau(x) \zeta(y) \tau(z), \tau(x) \tau(y) \zeta(z)); \\
& \mathrm{w}_3(\mathbf{r}) = \tau(x) \tau(y) \tau(z); 
\end{flalign}
\end{subequations}
and

\begin{equation}
\label{eq9}
\Phi^{(3)} = -n_0 \log \bigl(1- n_3 \bigr) + \frac{\mathbf{n_1}\cdot\mathbf{n_2}}{1- n_3} + \frac{n_2^{(x)} n_2^{(y)} n_2^{(z)}}{(1- n_3)^2}.
\end{equation}

\subsection{Functional minimization}

Once an explicit expression for the functional dependence of $\mathcal{F}^{\mathrm{exc}}$ on the single-particle density $\rho(\mathbf{r})$ is established, the equilibrium Helmholtz free energy $F(T,V,N)$ of $N$ particles at temperature $T$ in a volume $V$ (area $A$ for $D=2$) is obtained as the minimum of Eq. (\ref{eq1}) with respect to $\rho(\mathbf{r})$ under the constraint that

\begin{equation}
\label{eq2}
\int d^D \mathbf{r} \, \rho(\mathbf{r}) = N.
\end{equation}
In the case of hard-core interactions between the particles, the system is athermal and its thermodynamic state is completely identified by the packing fraction $\eta=N v_p / V$, where $v_p=\sigma^3$ is the particle's volume (for $D=2$ dimensions $\eta=N a_p / A$ and $a_p=\sigma^2$).

The numerically easiest way to solve the functional minimization problem consists of expressing the single-particle density in terms of a limited number of variational parameters. After inserting this ansatz into the free energy, the latter is minimized with respect to the variational parameters to obtain an estimate of the free energy at equilibrium. This approach has been widely applied in studying the freezing transition of hard spheres, where the single-particle density was parameterized as a sum of Gaussian functions centered on the lattice sites of the expected stable crystal phase ({\it Gaussian parameterization} or {\it ansatz}) \cite{tarazona, rosenfeld2, roth2}. In this paper we investigate the freezing transition of squares and cubes into square and simple-cubic crystal phases, for which the Gaussian ansatz can be expressed as

\begin{equation}
\label{eq10}
\rho_{\gamma, \lambda} (\mathbf{r}) = \eta \Bigl( \frac{\lambda}{\sigma} \Bigr) ^D \Bigl( \frac{\gamma}{\pi} \Bigr)^{\frac{D}{2}} \sum_{\mathbf{n} \in \mathbb{Z}^D} \exp \biggl [ - \gamma (\mathbf{r}-\lambda \mathbf{n})^2 \biggr].
\end{equation}
In Eq. (\ref{eq10}) the variational parameters are half of the inverse variance $\gamma$ and the lattice constant $\lambda$. Note that the lattice constant $\lambda$ is related to the vacancy concentration of the crystal $x_{vac}=(N_{\rm sites}-N)/N_{\rm sites}$ through $x_{vac}=1-\eta (\lambda/\sigma)^D$, where $N_{\rm sites}$ is the total number of sites.

An alternative to the Gaussian parameterization consists of the numerical solution of the Euler-Lagrange equation associated with the minimization problem ({\it free minimization}) \cite{oettel,roth}. By performing a functional differentiation of Eq. (\ref{eq1}) and imposing the constraint of Eq. (\ref{eq2}), one finds that the equilibrium $\rho(\mathbf{r})$ satisfies the following self-consistency equation

\begin{equation}
\label{eq11}
\rho(\mathbf{r}) = N \exp \Bigl[- \frac{\delta \beta \mathcal{F}^{\mathrm{exc}}}{\delta \rho(\mathbf{r})} \Bigr] \Biggl \{ \int d^D \mathbf{r}' \, \exp \Bigl[- \frac{\delta \beta \mathcal{F}^{\mathrm{exc}}}{\delta \rho(\mathbf{r}')} \Bigr] \Biggr \}^{-1}.
\end{equation}
At high packing fraction $\eta$ one expects the free energy to be minimized by inhomogeneous solutions characterized by spatial modulations of $\rho(\mathbf{r})$ along one or more directions. In practice, these spatial modulations must be inserted explicitly into Eq. (\ref{eq11}) by means of a Fourier series expansion. Therefore, the single particle density of a phase characterized by a $d$-dimensional spontaneous breaking of the translational symmetry is obtained by solving the following equation   

\begin{equation}
\label{eq12}
\rho(\mathbf{s}) = \eta \frac{\lambda^d}{\sigma^D} \exp \Bigl[- \frac{\delta \beta \mathcal{F}^{\mathrm{exc}}}{\delta \rho(\mathbf{s})} \Bigr] \Biggl \{ \int_{\Gamma} d^d \mathbf{s}' \, \exp \Bigl[- \frac{\delta \beta \mathcal{F}^{\mathrm{exc}}}{\delta \rho(\mathbf{s}')} \Bigr] \Biggr \}^{-1},
\end{equation}
where $\mathbf{s} \in \Gamma$ is a $d$-dimensional vector, $\Gamma=[-\lambda/2,\lambda/2]^d$ and $\lambda$ is the periodicity of the inhomogeneous solution (assumed to be the same along all the $d$ directions). The minimization procedure consists of (i) solving Eq. (\ref{eq12}) at fixed $\lambda$ and (ii) identifying the value of $\lambda$ that minimizes the free energy Eq. (\ref{eq1}). For hard squares  ($D=2$) we will see that Eq. (\ref{eq12}) describes smectic ($Sm$, $d=1$) and square crystal ($X$, $d=2$) phases; for hard cubes ($D=3$) Eq. (\ref{eq12}) accounts for smectic ($Sm$, $d=1$), columnar ($Col$, $d=2$) and simple-cubic crystal ($X$, $d=3$) ordering.

A numerical solution of Eq. (\ref{eq12}) on a grid of points is expected to offer a better description of the single particle density, and therefore a lower minimum free energy, than by minimizing the free energy by means of the Gaussian ansatz Eq. (\ref{eq10}). We develop a Picard algorithm to solve Eq. (\ref{eq12}), where all the convolutions involved in the FMT formalism are handled by means of Fast Fourier Transforms \cite{fft}. Moreover, the minimization with respect to the lattice spacing $\lambda$ is performed using the Brent algorithm \cite{brent}.

\section{Results}

\subsection{Parallel Hard Squares (D=2)}

\begin{figure}
%\center
\includegraphics[scale=0.67]{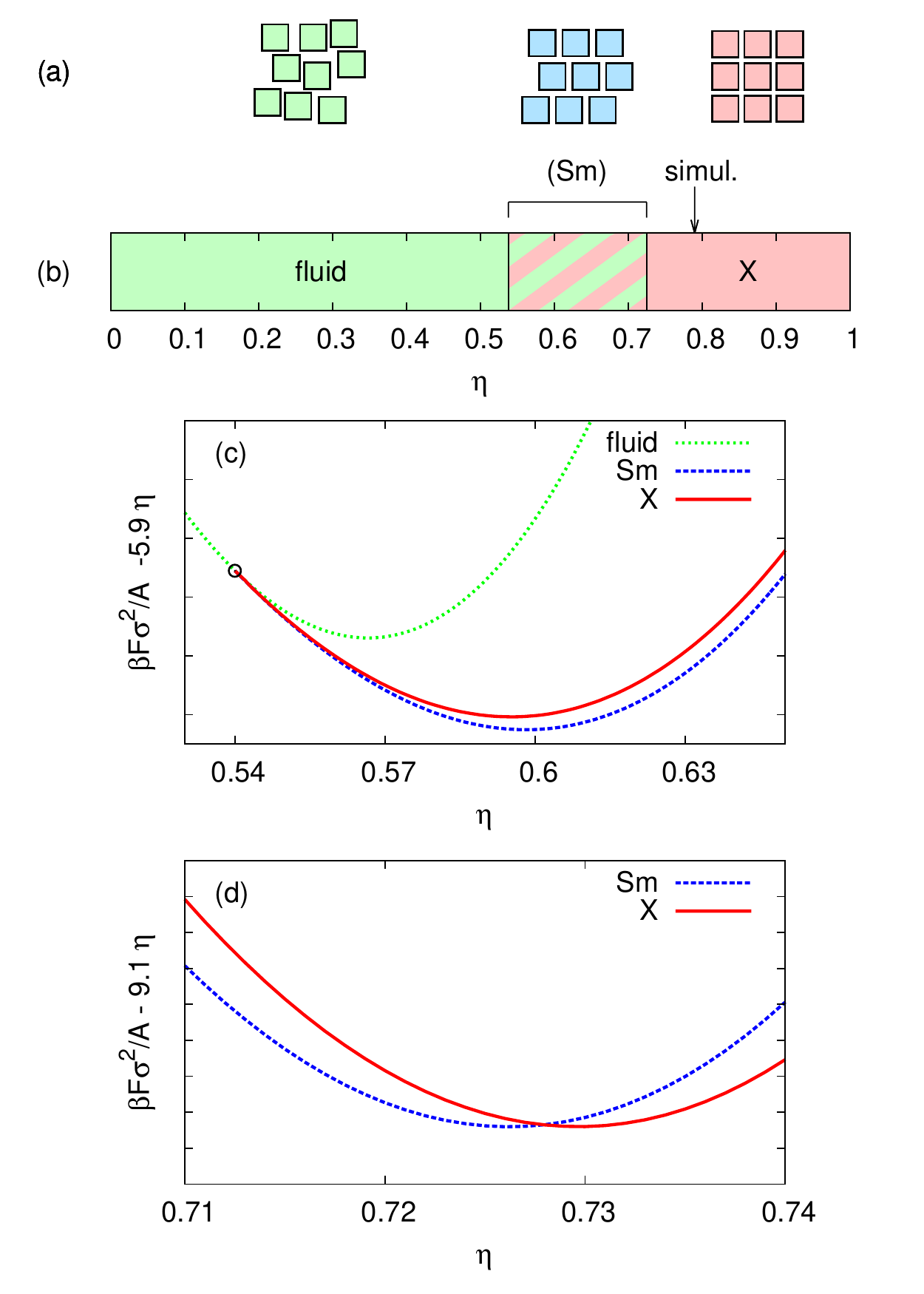}
\caption{\label{fig1} (a) Pictorial representation of (from left to right) the fluid, smectic ($Sm$) and square crystal ($X$) phases of parallel hard squares. (b) Phase diagram of parallel hard squares according to FMT, to be compared with the simulation value of Ref. \cite{hoover2} for the fluid-to-crystal transition packing fraction (vertical arrow). The (Sm) interval highlights the states where the (Peierls-Landau unstable) smectic phase is predicted to be the stable phase. (c) FMT prediction for the free energy per unity area $F/A$ of fluid (dotted green), $Sm$ (dashed blue) and $X$ (solid red lines) phases in the proximity of the second-order fluid-to-smectic (black circle) and (d) the first-order smectic-to-crystal transition.} 
\end{figure}

When considering the high-density phase behavior, monodisperse squares (as well as cubes in $D=3$ dimensions) possess a peculiar property. Unlike other regular polygons (e.g., regular triangles, pentagons, hexagons,..., and disks), squares do not have a well-defined ``locked-in'' configuration at close packing. In other words, besides the two-dimensional ordered square crystal ($X$), any other configuration with rows (or columns) shifted with respect to one another completely fills the plane. Therefore, also a smectic phase ($Sm$), characterized by positional ordering along one direction only, should in principle be considered as a candidate stable phase (see Fig. \ref{fig1}(a)). The higher degeneracy of $Sm$ configurations with respect to $X$ configurations suggests a higher entropy of the former with respect to the latter. However, in low-dimensional systems thermal fluctuations from equilibrium can play a relevant role in destroying long-range order, leading to so-called Landau-Peierls instabilities \cite{
peierls}. In particular, for short-range interactions proper crystals do not exist in $D=2$ dimensions, since positional ordering can in this case have at most quasi-long-range character \cite{mermin}. The situation is even more dramatic when considering smectic phases, where thermal fluctuations make the correlation between layers decay exponentially with the distance \cite{toner}. This means that in $D=2$ dimensions we do not expect smectic ordering to be stable at all. Computer simulations of both parallel \cite{ree} and freely-rotating \cite{wojc} hard squares, where only a direct fluid-to-crystal phase transition was observed without any smectic state, confirm this picture.

We report in Fig. \ref{fig1}(b) the phase diagram of parallel hard squares, as obtained by freely minimizing the FMT functional with respect to the single-particle density $\rho(\mathbf{r})$ including vacancies. Despite the above-mentioned considerations on the effect of fluctuations, we approximate the single-particle density of the $X$ phase by assuming long-range order (cf. Eqs. (\ref{eq10}) and (\ref{eq12})). A similar approximation was recently applied for the description of the freezing transition in two-dimensional hard disks, showing remarkably good agreement with computer simulations \cite{roth}. We include in our calculations also the possibility of long-range $Sm$ ordering, which is however expected to be Landau-Peierls unstable. The free-energy dependence of the fluid, smectic and crystal phases on the packing fraction is reported in Fig. \ref{fig1}(c) and (d) for two density intervals. Note that this representation allows for common tangent construction to identify coexisting states. Surprisingly, FMT 
predicts a second-order fluid-to-smectic transition at $\eta^{*}=0.538$ (Fig. \ref{fig1}(c)) and a weakly first-order smectic-to-crystal transition with bulk coexisting densities $\eta_{Sm}=0.726$ and $\eta_{X}=0.730$ (Fig. \ref{fig1}(d)). The picture does not change appreciably by minimizing the free energy within the Gaussian ansatz, giving the sole effect of slightly displacing the alleged $Sm-X$ transition ($\eta_{Sm}=0.750$ and $\eta_{X}=0.756$, not shown). As already pointed out, theoretical considerations and simulation results rule out the possibility of stable smectic ordering in the thermodynamic limit. Therefore, we must conclude that the smectic phase is an artifact due to the mean-field character of the Fundamental Measure Theory, which is unable to take fully into account the role of long-wavelength fluctuations from equilibrium. On the other hand, the question whether the fluid or the crystal is the stable phase in the range of allegedly smectic stability (striped region in Fig. \ref{fig1}(b)) is open. We address this point, as well as possible conditions of smectic stability, in the final discussion of Sec. \ref{conclusions}.

\begin{figure}
%\center
\includegraphics[scale=0.67]{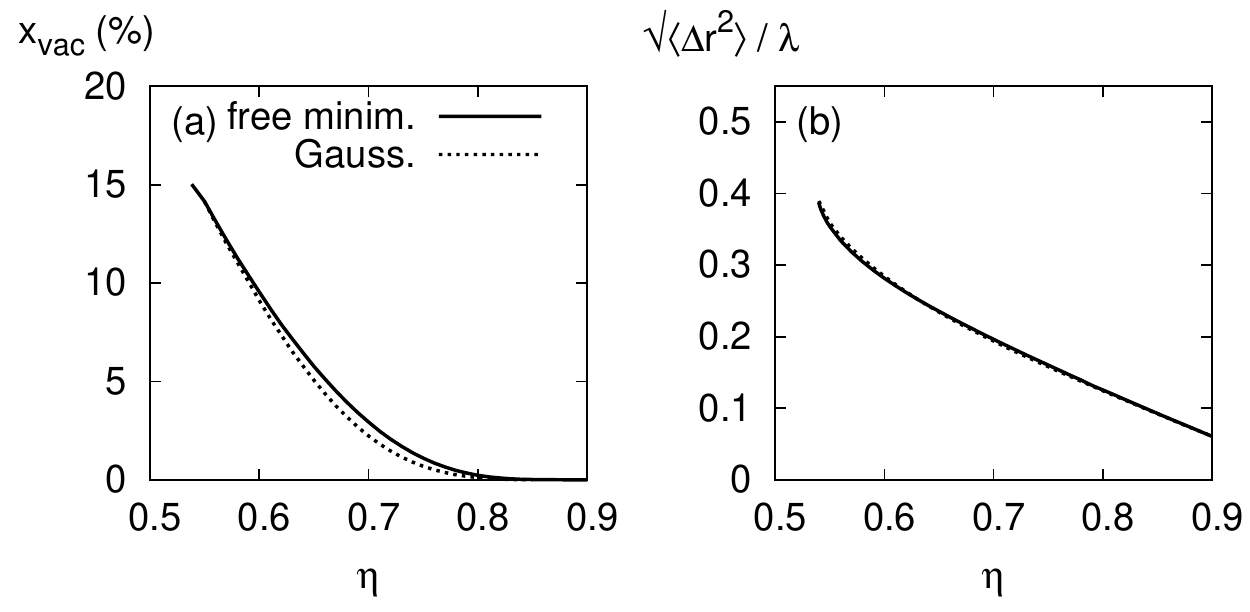}
\caption{\label{fig2} FMT results for (a) the vacancy concentration and (b) the root-mean-squared deviation from the average position (in units of the lattice constant $\lambda$) of the square crystal phase of parallel hard squares. Solid lines correspond to values calculated by free minimization of the FMT functional, whereas dashed lines indicate those obtained through the Gaussian ansatz.} 
\end{figure}   

In order to further investigate the properties of the crystal, we report the dependence of the vacancy concentration (Fig. \ref{fig2}(a)) and the root-mean-squared deviation from the average position in the unit cell, also known as Lindemann parameter (Fig. \ref{fig2}(b)), on the packing fraction. At the  second-order transition at $\eta=0.538$ the vacancy concentration is $x_{vac}\simeq15\%$, and $x_{vac}$ reduces monotonically with $\eta$. This value is appreciably higher than that predicted for hard disks $x_{vac}\simeq2\%$ \cite{roth}, in analogy to the $D=3$ dimensional case of hard-cubes and hard-spheres systems, where the vacancy concentration at the fluid-coexistence in the former is two orders of magnitude higher than in the latter \cite{smallenburg,marechal,oettel}. Not unlike the case of hard cubes \cite{smallenburg}, these results suggest that vacancies can play an important role in stabilizing (quasi-long-range) crystal order in systems of hard squares. Moreover, Fig. \ref{fig2}(a) highlights a 
systematic, albeit small, underestimation of the vacancy concentration at intermediate packing fraction when the Gaussian ansatz is applied. On the other hand, no appreciable difference with the root-mean-squared deviation calculated by free minimization is observed in Fig. \ref{fig2}(b). In both cases, at the second-order transition the Lindemann parameter assumes a value close to $0.4$, remarkably higher than that of $0.15$ expected at melting for three-dimensional systems \cite{hansen}, and decays to zero towards close packing.

\begin{figure}
%\center
\includegraphics[scale=0.67]{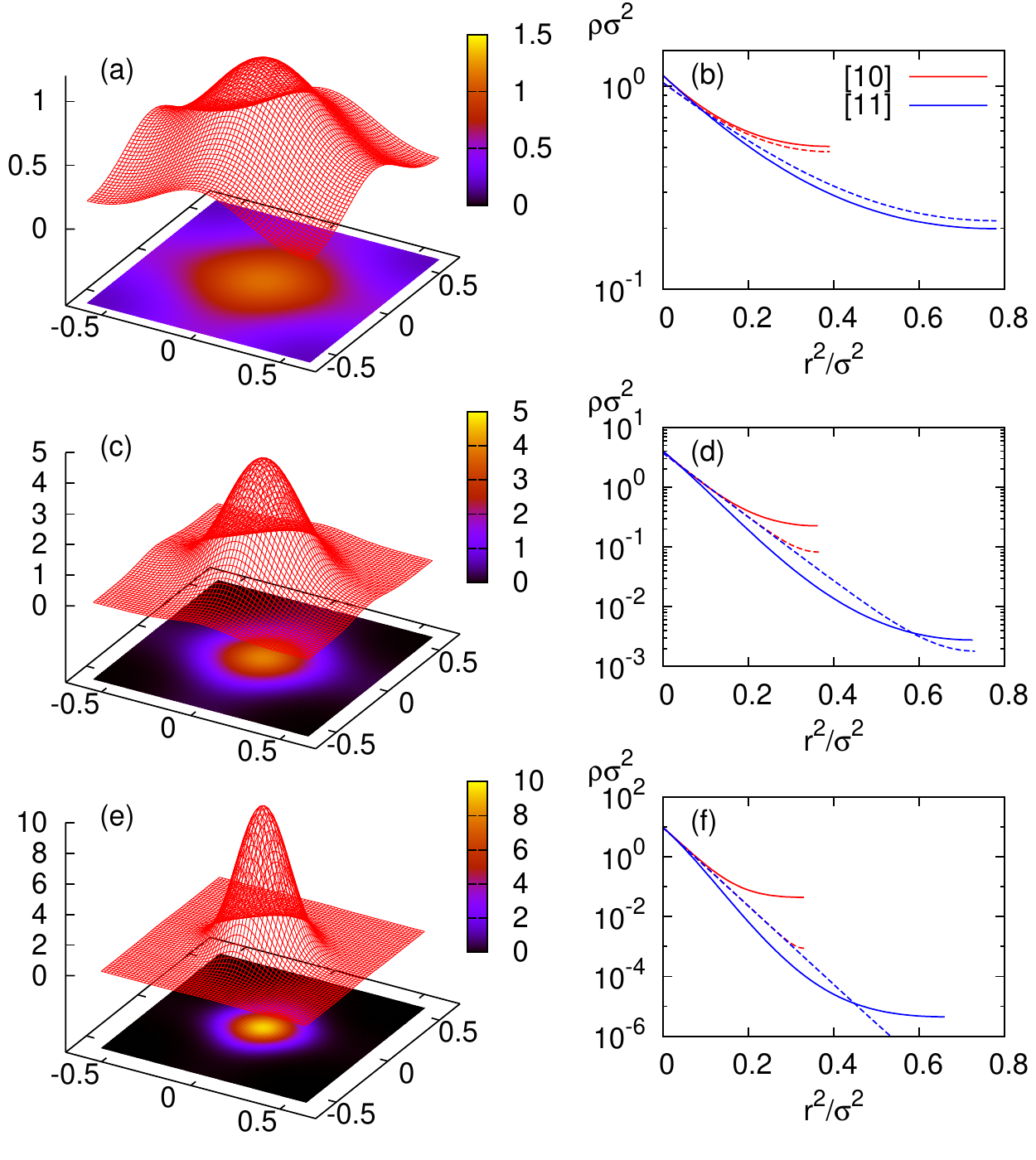}
\caption{\label{fig3} Single-particle density $\rho(x,y)$ in the unit cell of the square crystal phase obtained through FMT free minimization at packing fraction (a) $\eta=0.55$, (c) $\eta=0.65$ and (e) $\eta=0.75$. On the right column we report the section of the single-particle density along the [10] (red lines) and [11] (blue lines) crystallographic directions. The graphs are expressed in log-scale as a function of the squared distance from the center of the unit cell. The dashed lines represent the corresponding functional dependence obtained through the Gaussian ansatz.}   
\end{figure}   

We complete our analysis of the crystal phase by studying the evolution of the single-particle density from the freezing-transition region, where $\rho(x,y)$ is still appreciably non-zero at the edge of the Wigner-Seitz cell, to the confined regime at higher density. In Fig. \ref{fig3}(a), (c) and (e) we report the functional dependence of the equilibrium single-particle density inside the unit cell at packing fraction $\eta=0.55$, $0.65$ and $0.75$, respectively. In order to ease the analysis, we plot on the right of each figure (Fig. \ref{fig3}(b), (d) and (f)) a section of the corresponding $\rho(x,y)$ along the crystallographic directions [10] and [11]. These graphs are represented in logarithmic scale as a function of $r^2$ to highlight Gaussian behavior (straight lines). As expected, the section along the [10] direction, connecting nearest-neighbor sites, is systematically bigger than that along the [11] direction for both the freely minimized (solid lines) and Gaussian-parameterized (dashed lines) 
profiles. At the lowest packing fraction the peak of the density distribution is smeared out on the unit cell. As a consequence, the single-particle density in the Gaussian parameterization shows relevant deviations from the Gaussian distribution due to the overlap of the peaks centered on neighboring cells. In this way, the tails of the neighboring lattice sites allow to properly account for the anisotropy of the density peak, thus leading to a marked difference between the [10] and [11] profiles, similar to the case of free minimization (Fig. \ref{fig3}(b)). This overlap is weaker at higher packing fraction, where the confinement is stronger; hence, the Gaussian ansatz fails to reproduce the anisotropy of the distribution in this regime (Fig. \ref{fig3}(d) and (f)). However, these deviations occur on a density scale a few orders of magnitude smaller than the peak value, and therefore their relevance is quantitatively limited.      

\subsection{Parallel Hard Cubes (D=3)}

Here we compare the predictions of FMT for parallel hard cubes, extensively studied in the past by means of the Gaussian parameterization \cite{mraton, groh, marechal}, with our results based on the free minimization of the functional. Since the formulation by Cuesta and Mart\'{\i}nez-Rat\'{o}n \cite{cuesta}, FMT is known to correctly predict two significant properties of the freezing transition of the model \cite{mraton}: (i) its second-order character and (ii) the role of vacancies in stabilizing the crystal. 

\begin{figure}
%\center
\includegraphics[scale=0.67]{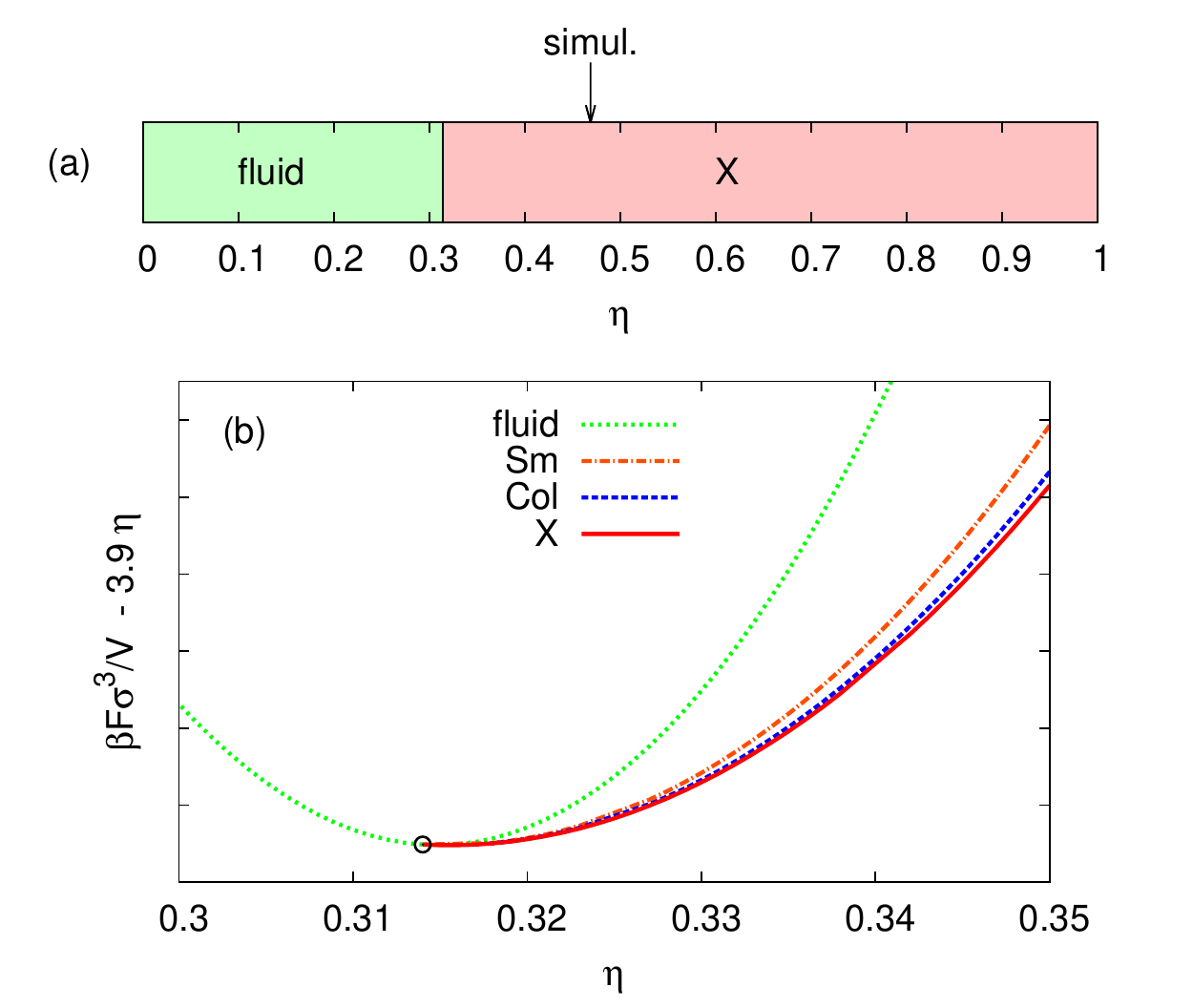}
\caption{\label{fig4} (a) Phase diagram of parallel hard cubes according to FMT, to be compared with the simulation result of Ref. \cite{marechal} for the fluid-to-crystal ($X$) transition packing fraction (vertical arrow). (b) FMT prediction for the free energy per unity volume $\beta F/V$ of fluid (dotted green), crystal (solid red) and the metastable columnar ($Col$, dashed blue) and smectic ($Sm$, dot-dashed orange) phases in the proximity of the second-order freezing transition (black circle).} 
\end{figure}

The second-order fluid-to-crystal transition, which is known to become first-order when the rotational degrees of freedom are taken into account \cite{jagla}, is predicted to occur at $\eta=0.314$. As in the case of parallel squares, this value appreciably underestimates the simulation result of $\eta=0.469$ \cite{marechal} (see Fig. \ref{fig4}(a)). Also in analogy with 
parallel hard squares, parallel cubes lack a ``locked-in'' configuration at close packing. By means of a bifurcation analysis of the FMT functional and computer simulations, Groh and Mulder addressed the question about the stability of columnar order, and showed it to be metastable. In contrast to hard squares, smectic and columnar solutions are in the three-dimensional case always metastable with respect to the crystal. This finding, which results directly from our free minimization scheme, is easily verified by comparing the free-energy curves of smectic, columnar and crystal phases as a function of the packing fraction in Fig. \ref{fig4}(b).

\begin{figure}
%\center
\includegraphics[scale=0.67]{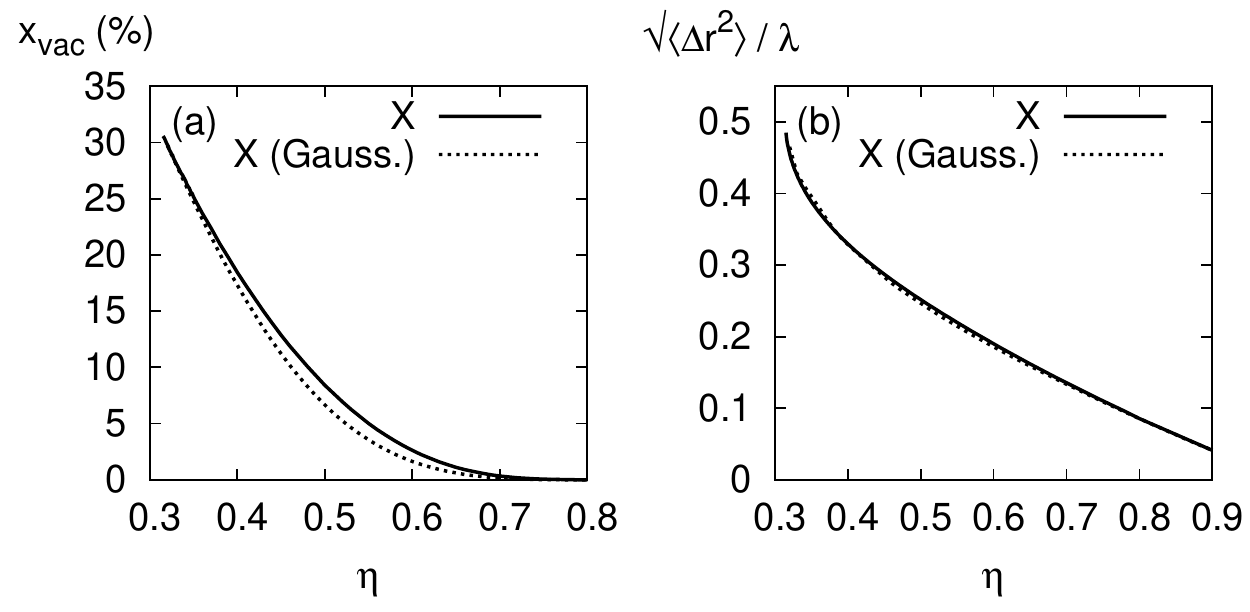}
\caption{\label{fig5} FMT results for (a) the vacancy concentration and (b) the root-mean-squared deviation from the average position (in units of the lattice constant $\lambda$) of the simple-cubic crystal phase of parallel hard cubes. Solid lines correspond to values calculated by free minimization of the FMT functional, whereas dashed lines indicate those obtained through the Gaussian ansatz.} 
\end{figure}  

The remarkably high concentration of vacancies at the freezing transition, $x_{vac} \simeq 30\%$, is a known feature of the theory (cf. Fig. \ref{fig5}(a)) \cite{mraton, groh}. Despite this value being three orders of magnitude higher than that measured for hard spheres \cite{oettel}, it was shown to be compatible with computer simulations of both parallel ($x_{vac}=13\%$ \cite{marechal}) and freely-rotating ($x_{vac}=6.4\%$ \cite{smallenburg}) hard cubes, thus highlighting the essential role of vacancies in stabilizing the simple-cubic crystal. Within the free minimization of the FMT functional, the vacancy concentration at bulk coexistence does not change with respect to the Gaussian ansatz result. Nonetheless, an inspection of Fig. \ref{fig5}(a), reporting $x_{vac}$ as a function of the packing fraction $\eta$, shows that the Gaussian ansatz tends to underestimate this property at intermediate packing fractions. Therefore, in this regime the free minimization improves the Gaussian ansatz data by 
furnishing results closer to those of computer simulation \cite{marechal}. However, if we focus on the root-mean-squared deviation from the average lattice site (Lindemann parameter), reported as a function of $\eta$ in Fig. \ref{fig5}(b), we do not observe any appreciable deviation from the known dependence calculated by means of the Gaussian ansatz.

\begin{figure}
%\center
\includegraphics[scale=0.67]{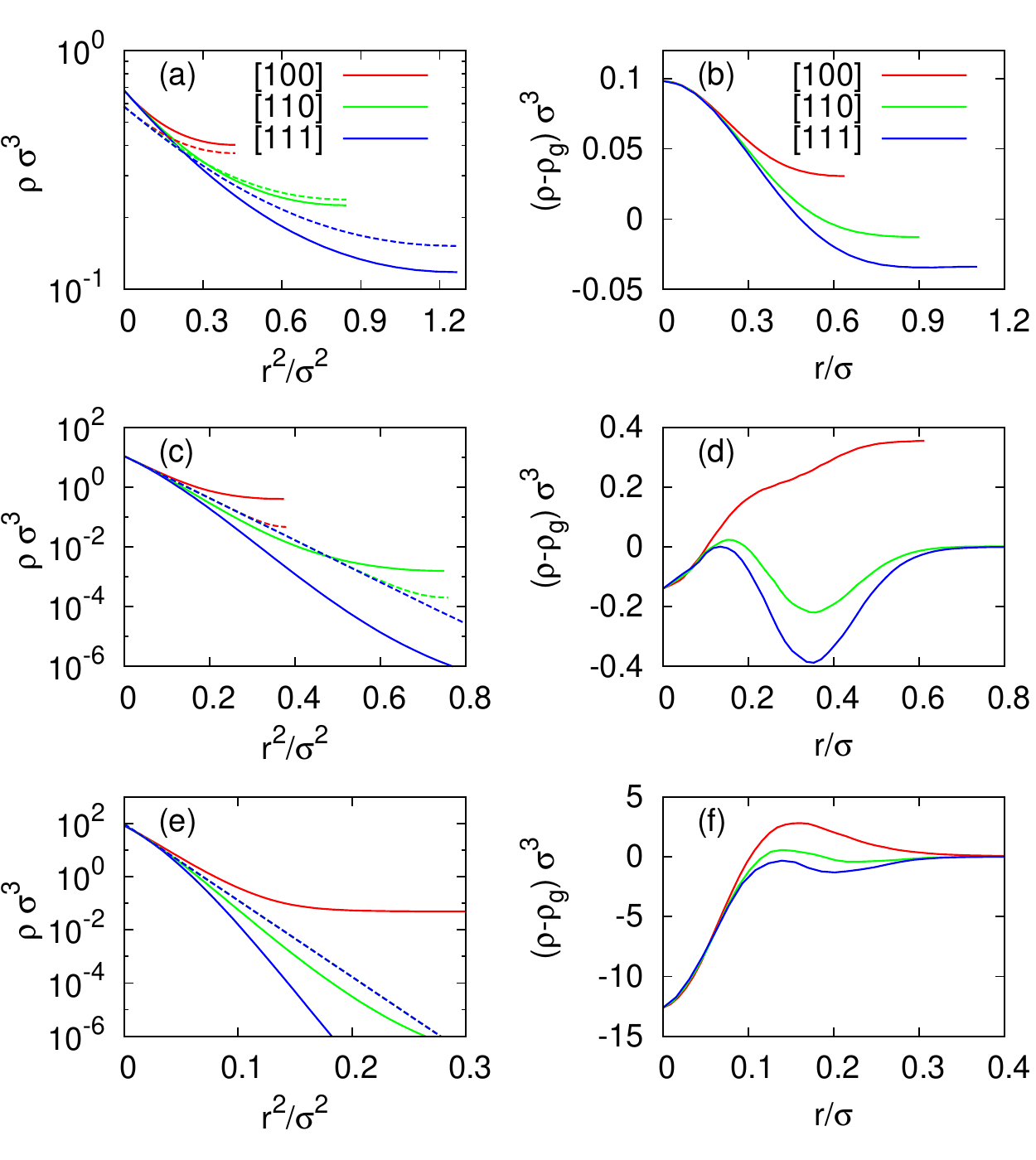}
\caption{\label{fig6} FMT prediction for the single-particle density of parallel hard cubes. The graphs on the left show sections of $\rho(x,y,z)$ along the crystallographic directions [100] (red lines), [110] (green lines) and [111] (blue lines) at packing fraction (a) $\eta=0.32$, (c) $0.50$ and (e) $0.70$ calculated by means of free minimization of the functional (solid lines) and the Gaussian ansatz (dashed lines). The right graphs represent the corresponding absolute difference between the free minimization solution and the Gaussian ansatz for the three packing fractions ((b)$\eta=0.32$, (d) $0.50$ and (f) $0.70$) and the three crystallographic directions considered.}   
\end{figure}    

Finally, in Fig. \ref{fig6} we represent sections of the equilibrium single-particle density $\rho(x,y,z)$ at packing fraction $\eta=0.32$ ((a)-(b)), $0.50$ ((c)-(d)) and $0.70$ ((e)-(f)) and we compare them with the corresponding Gaussian ansatz solution (dashed lines). The graphs on the left ((a), (c) and (e)) show sections along the crystallographic directions [100] (red lines), [110] (green lines) and [111] (blue lines); to ease the comparison, we report on the right of each graph ((b), (d) and (f)) the absolute difference of these sections between the two minimization methods. In the three cases, the cubic symmetry of the freely-minimized solution is evident by the hierarchy in values of the single particle density along the three crystallographic directions. In analogy with the parallel square system, at low enough packing fraction there is good quantitative agreement between the two methods, since the overlap between neighboring peaks within the Gaussian ansatz
allows to reproduce the anisotropy of the single-particle distribution. At higher packing fraction, deviations from the Gaussian-ansatz solution are evident, but limited to the low-density region of the unit cell.

\section{Discussion and Conclusions}
\label{conclusions}

By means of Fundamental Measure Theory we investigate the phase behavior of single-component systems of parallel hard squares (in $D=2$ dimensions) and cubes ($D=3$). Our attention focuses on the predictions for the freezing transition and the properties of the crystal phase. In density-functional theory the typical approach for describing crystal phases is based on the parameterization of the single-particle density by a sum of Gaussian functions centered on the lattice sites. We compare these predictions with a more accurate free-minimization method, where the single-particle density is evaluated on a grid of points.

Despite its simplicity, we conclude that for both squares and cubes the Gaussian parameterization works remarkably well. Apart from some inadequacy of the Gaussian ansatz in describing the anisotropy of the single-particle density of the crystal, the main deviations between the two minimization methods lie in the expected vacancy concentrations of the square and simple-cubic crystals, which appears to be slightly underestimated by the Gaussian ansatz. On the other hand, as already noticed for cubes, FMT suffers from a serious inability to give quantitatively reliable values for the freezing packing fraction. However, improvement in this direction can be achieved only by a reformulation of the theory itself, as the numerical minimization is performed exactly.

For the three-dimensional system of parallel hard cubes our results coincide with previous FMT analysis based on the Gaussian parameterization and indicate a second-order vacancy-rich fluid-to-crystal transition. For the parallel hard-square system, this work constitutes to the best of our knowledge the first analysis based on density-functional theory. In contrast with previous simulation studies, Fundamental Measure Theory predicts a stable smectic phase in between the low-density fluid and the high-density square crystal. However, by taking into account the effect of long-wavelength thermal fluctuations, one can show the one-dimensional smectic ordering to be unstable. Therefore, we deduce that the mean-field character of the theory, which is unable to properly take into account the role of fluctuations from equilibirum, is the element to be blamed for this artifact. When big enough simulation boxes are considered, computer simulations with periodic boundary conditions comfirm the picture of an unstable smectic phase \cite{wojc}. However, it is interesting to notice that, when the simulation box is small enough, ordering of the squares in parallel layers was observed \cite{wojc}. On the basis of these observations, it is tempting to conclude that, when long-wavelength fluctuations can be neglected, the behavior of the system coincides with the predictions of FMT, showing a stable smectic phase. In other words, we expect parallel-squares systems to develop intermediate smectic states in finite size systems and under the effect of confining walls.

For what concerns the original problem regarding the phase behavior of parallel hard squares in the thermodynamic limit, the conclusions we can draw are more limited. In fact, on the basis of our theoretical results we do not have enough elements to deduce which of the two phases, either the fluid or the crystal, is the stable one in the density range where the theory predicts a stable smectic. Therefore, we cannot conclude whether the theory predicts a low-density second-order freezing transition with a high vacancy concentration in the crystal, or a higher-density weakly first-order freezing with a lower vacancy concentration. Addressing this problem is challenging, as it should involve a proper incorporation of long-wavelength Landau-Peierls fluctuations into the 
density-functional theory. 

\section*{Acknowledgements}

It is a pleasure to thank Prof. R. Evans for a critical reading of the manuscript and F. Smallenburg and M. Marechal for stimulating discussions. 

This work is financed by an NWO-VICI grant and is part of the research program of the ``Stichting voor Fundamenteel Onderzoek der Materie (FOM)'', which is financially supported by the ``Nederlandse Organisatie voor Wetenschappelijk Onderzoek (NWO)''.

\end{document}